\documentclass[aps,prl,preprint,tightenlines,superscriptaddress,showpacs,byrevtex]{revtex4}

\newcommand{\bm}[1]{\mbox{\boldmath $#1$}}

\usepackage{graphicx}
\usepackage{color}

\begin{document}

% --------------------------------------------------

\preprint{\vbox{ \hbox{   }
                 \hbox{BELLE-CONF-0433}
                 \hbox{ICHEP04 12-0680} 
}}

\title{ \quad\\[0.5cm] Search for Lepton and Baryon Number Violating $\tau^-$ Decays
into $\overline{p}\gamma$, $\overline{p}\pi^0$,
$\overline{\Lambda}\pi^-$, and $\Lambda \pi^-$}

%%%% >>>>> insert the authorlist here. BEFORE the abstract !!!!! <<<<<

%%% Paper:    
%%% Journal:  summer 2004 conference papers (PRL format)
%%% Contacts: 
%%% Last revised on July 14, 2004 16:40:00 EDT
%%% Non-responding authors or those who said NO are commented out.
%%% ====================================================================
%%% Click the RELOAD button on your web browser to see the updated file.
%%% ====================================================================
%%% Use \input{author} to insert this material into your latex file.
%%%%% Force institutions to appear in alphabetical order when typeset.
\affiliation{Aomori University, Aomori}
\affiliation{Budker Institute of Nuclear Physics, Novosibirsk}
\affiliation{Chiba University, Chiba}
\affiliation{Chonnam National University, Kwangju}
\affiliation{Chuo University, Tokyo}
\affiliation{University of Cincinnati, Cincinnati, Ohio 45221}
\affiliation{University of Frankfurt, Frankfurt}
\affiliation{Gyeongsang National University, Chinju}
\affiliation{University of Hawaii, Honolulu, Hawaii 96822}
\affiliation{High Energy Accelerator Research Organization (KEK), Tsukuba}
\affiliation{Hiroshima Institute of Technology, Hiroshima}
\affiliation{Institute of High Energy Physics, Chinese Academy of Sciences, Beijing}
\affiliation{Institute of High Energy Physics, Vienna}
\affiliation{Institute for Theoretical and Experimental Physics, Moscow}
\affiliation{J. Stefan Institute, Ljubljana}
\affiliation{Kanagawa University, Yokohama}
\affiliation{Korea University, Seoul}
\affiliation{Kyoto University, Kyoto}
\affiliation{Kyungpook National University, Taegu}
\affiliation{Swiss Federal Institute of Technology of Lausanne, EPFL, Lausanne}
\affiliation{University of Ljubljana, Ljubljana}
\affiliation{University of Maribor, Maribor}
\affiliation{University of Melbourne, Victoria}
\affiliation{Nagoya University, Nagoya}
\affiliation{Nara Women's University, Nara}
\affiliation{National Central University, Chung-li}
\affiliation{National Kaohsiung Normal University, Kaohsiung}
\affiliation{National United University, Miao Li}
\affiliation{Department of Physics, National Taiwan University, Taipei}
\affiliation{H. Niewodniczanski Institute of Nuclear Physics, Krakow}
\affiliation{Nihon Dental College, Niigata}
\affiliation{Niigata University, Niigata}
\affiliation{Osaka City University, Osaka}
\affiliation{Osaka University, Osaka}
\affiliation{Panjab University, Chandigarh}
\affiliation{Peking University, Beijing}
\affiliation{Princeton University, Princeton, New Jersey 08545}
\affiliation{RIKEN BNL Research Center, Upton, New York 11973}
\affiliation{Saga University, Saga}
\affiliation{University of Science and Technology of China, Hefei}
\affiliation{Seoul National University, Seoul}
\affiliation{Sungkyunkwan University, Suwon}
\affiliation{University of Sydney, Sydney NSW}
\affiliation{Tata Institute of Fundamental Research, Bombay}
\affiliation{Toho University, Funabashi}
\affiliation{Tohoku Gakuin University, Tagajo}
\affiliation{Tohoku University, Sendai}
\affiliation{Department of Physics, University of Tokyo, Tokyo}
\affiliation{Tokyo Institute of Technology, Tokyo}
\affiliation{Tokyo Metropolitan University, Tokyo}
\affiliation{Tokyo University of Agriculture and Technology, Tokyo}
\affiliation{Toyama National College of Maritime Technology, Toyama}
\affiliation{University of Tsukuba, Tsukuba}
\affiliation{Utkal University, Bhubaneswer}
\affiliation{Virginia Polytechnic Institute and State University, Blacksburg, Virginia 24061}
\affiliation{Yonsei University, Seoul}
  \author{K.~Abe}\affiliation{High Energy Accelerator Research Organization (KEK), Tsukuba} % KEK
  \author{K.~Abe}\affiliation{Tohoku Gakuin University, Tagajo} % TohokuGakuin
  \author{N.~Abe}\affiliation{Tokyo Institute of Technology, Tokyo} % TIT
  \author{I.~Adachi}\affiliation{High Energy Accelerator Research Organization (KEK), Tsukuba} % KEK
  \author{H.~Aihara}\affiliation{Department of Physics, University of Tokyo, Tokyo} % Tokyo
  \author{M.~Akatsu}\affiliation{Nagoya University, Nagoya} % Nagoya
  \author{Y.~Asano}\affiliation{University of Tsukuba, Tsukuba} % Tsukuba
  \author{T.~Aso}\affiliation{Toyama National College of Maritime Technology, Toyama} % Toyama
  \author{V.~Aulchenko}\affiliation{Budker Institute of Nuclear Physics, Novosibirsk} % BINP
  \author{T.~Aushev}\affiliation{Institute for Theoretical and Experimental Physics, Moscow} % ITEP
% ...
% Lausanne
\collaboration{The Belle Collaboration}

%---------- End of author list ----------
%\collaboration{Belle Collaboration}

\noaffiliation

\begin{abstract}
We have searched for decays that violate
both lepton and baryon number using data collected 
by the Belle detector at the KEKB asymmetric $e^+e^-$ collider.
No signals are found in 
$\tau^- \rightarrow \overline{p} \gamma$, 
$\tau^- \rightarrow \overline{p} \pi^0$, 
$\tau^- \rightarrow \overline{\Lambda} \pi^-$, 
and
$\tau^- \rightarrow \Lambda \pi^-$ 
and we set upper limits on the branching fractions of
${\cal{B}}(\tau^-\rightarrow \overline{p}\gamma) < 3.0\times 10^{-7}$, 
${\cal{B}}(\tau^-\rightarrow \overline{p}\pi^0) < 6.5\times 10^{-7}$, 
${\cal{B}}(\tau^-\rightarrow \overline{\Lambda}\pi^-) < 1.3\times 10^{-7}$,
and 
${\cal{B}}(\tau^-\rightarrow \Lambda \pi^-) < 0.70\times 10^{-7}$ 
at the 90\% confidence level. 
The former two results improve the previous limits by a factor 
of 12 and 23, respectively, 
while the latter two are the first searches ever performed. 
\end{abstract}


\maketitle

\section{Introduction}

While the Standard Model (SM) assumes
both the baryon number ($B$) and lepton number ($L$) conservation, 
in some extensions beyond the SM such as Grand Unified Theories(GUTs), 
$B$ and $L$ violation is expected 
while their difference $B-L$ is conserved~\cite{cite:GUT}.
In Ref.~\cite{cite:Hou:2004uc}, 
right-handed four-fermion couplings that conserve 
$B-L$ were used to consider $B$ and $L$ violating 
$\tau$ lepton, $D$ and $B$ meson decays.
High luminosity B-factories provide an opportunity to look for
such decays
with unprecedented sensitivity. 

We report here our searches for
$\tau^-\rightarrow \overline{p}\gamma$, 
$\overline{p}\pi^0$,
$\overline{\Lambda}\pi^-$, and $\Lambda\pi^-$ decays
with data samples of 86.7 fb$^{-1}$ for $\overline{p} \gamma$
and 153.8 fb$^{-1}$ for all other modes, 
collected with the Belle detector at the KEKB 
asymmetric $e^+e^-$ collider~\cite{kekb}. 
Previously, the best upper limits for the corresponding branching 
fractions were obtained by CLEO~\cite{cite:cleo}
based on a data sample of  4.7 fb$^{-1}$:
${\cal{B}}(\tau^-\rightarrow \overline{p}\gamma) < 3.5\times 10^{-6}$ and 
${\cal{B}}(\tau^-\rightarrow \overline{p}\pi^0) < 1.5\times 10^{-5}$ 
at 90\% C.L. The decay modes
%${\cal{B}}( \tau^-\rightarrow \overline{\Lambda}\pi^-$)
%and
%${\cal{B}}( \tau^-\rightarrow \Lambda\pi^-$)
$\tau^-\rightarrow \overline{\Lambda}\pi^-$
and
$\tau^-\rightarrow \Lambda\pi^-$
have never been studied before.
Unless otherwise stated, charge conjugate decays are implied throughout
this paper.

The Belle detector is a large-solid-angle magnetic spectrometer that
consists of a silicon vertex detector (SVD), 
a 50-layer central drift chamber (CDC), 
an array of aerogel threshold \v{C}erenkov counters (ACC), a barrel-like arrangement of 
time-of-flight scintillation counters (TOF), and an electromagnetic calorimeter 
comprised of CsI(Tl) crystals (ECL) located inside
a superconducting solenoid coil
that provides a 1.5~T magnetic field.  
An iron flux-return located outside of the coil is instrumented to detect $K_L^0$ mesons 
and to identify muons (KLM).  
The detector is described in detail elsewhere~\cite{Belle}.

We search for $\tau^+\tau^-$ events
in which one $\tau$ decays
into the mode studied (signal side) 
and the other $\tau$ (tag side) decays into one charged particle, 
photons and neutrino(s). 
The selection criteria are determined by examining Monte Carlo (MC) 
simulations for 
signal $\tau$-pair decays and
background (BG) events coming from 
generic $\tau$-pair decays ($\tau^+\tau^-$), $q\overline{q}$ continuum, $B\bar{B}$, 
Bhabha and $\mu\mu$ as well as two-photon processes. 
The KORALB/TAUOLA~\cite{cite:koralb_tauola} and QQ~\cite{cite:qq} 
generators are used for event generation, and
GEANT3~\cite{cite:geant3} 
is used to simulate the Belle detector response.
The two-body decays of 
the signal $\tau$ are assumed to have a uniform angular distribution
in the $\tau$'s rest frame. 
All the kinematical variables are calculated in the laboratory frame, 
while those in the $e^+ e^-$ center-of-mass (CM) frame are 
indicated by the superscript ``CM''. 

\section{Analysis of $\bm{\tau^-\rightarrow \overline{p}\gamma}$ and $\bm{\overline{p}\pi^0}$} 

\subsection{Event selection}

The experimental signature of these events 
has one $\tau$ lepton decaying into a proton and photons
(signal side) and the other decaying via the 1-prong mode (tag side):
\begin{equation}
\left\{ \tau^-\rightarrow \overline{p} + n_{\gamma}^{\rm{SIG}} \right\} + 
\left\{ \tau^+\rightarrow {\rm(a ~track)}^+ + n_{\gamma}^{\rm{TAG}} + X(\rm{missing}) \right\}, 
\end{equation}
where a track should have transverse momentum $p_t > 0.1$ GeV/$c$ 
and polar angle $-0.819 < \cos\theta < 0.906$.
The two highlighted tracks should have zero net charge. 
A proton track forming the signal 
$\overline{p}\gamma/ \overline{p}\pi^0$ is 
required to have $p > 1.0$ GeV/$c$ for reliable identification. 
A photon should have energy $E_{\gamma} > 0.1$ GeV in the same 
polar angle range as that for tracks. 
The number of photons on the signal side is
$1 < (n_{\gamma}^{\rm{SIG}} + n_{\gamma}^{\rm{TAG}}) < 3$ for 
$\tau^-\rightarrow \overline{p}\gamma$
and
$2 < (n_{\gamma}^{\rm{SIG}}+n_{\gamma}^{\rm{TAG}})< 4$ for 
$\tau^-\rightarrow \overline{p}\pi^0$. 

We divide the event into two hemispheres in the CM frame by 
the plane perpendicular to the thrust axis. 
The hemisphere containing a proton track (its identification is later explained), 
is the 'signal side', 
the opposite hemisphere is the 'tag side'.
We require an invariant mass of the visible particles on the tag side of
$M_{\rm{TAG}} < $ 1.2 GeV/$c^2$ for the $\overline{p} \gamma$ mode and 
$M_{\rm{TAG}} < m_{\tau}$ for the $\overline{p} \pi^0$ one to reduce 
background (BG) from the $e^+ e^- \rightarrow q\overline{q}$ continuum
process (where $q = u,\ d,\ s,\ c$). 

For $\overline{p}\gamma$,
we require the photon to have $E_{\gamma} > 0.5$ GeV.
In order to reduce the number of fake $\gamma$'s originating from 
a $\overline{n}$ in the $q\overline{q}$ 
continuum, the ratio of photon energy deposition in a $3\times 3$ ECL 
square to a $5\times 5$ square, $E9/E25$, is required to be 
larger than 0.93 since the hadronic shower of a $\overline{n}$
in the ECL is wider than a photon's electromagnetic shower.

The $\pi^0$ from $\overline{p}\pi^0$ is reconstructed from a $\gamma\gamma$
pair with an invariant mass 
within $\pm 5\sigma_{\pi^0}$ ($\sigma_{\pi^0} = 5-8$ MeV/$c^2$) of
the nominal $\pi^0$ mass. 
We impose a $\pi^0$ veto on photon(s) for
$\overline{p}\gamma/ \overline{p}\pi^0$ candidates: 
it should not reconstruct a $\pi^0$ meson when combined with
any other photon whose energy exceeds 50 MeV.

Correlations are considered among the tracks, photons and a missing particle that 
carries away undetected momentum and energy.
A requirement on the total visible energy 
$0.5 < E_{vis}^{\rm{CM}}/\sqrt{s} < 0.92$ 
is imposed to reject Bhabha scattering and $\mu^+\mu^-$ production. 
Restrictions on the opening angle between $p$ and $\gamma$, and $p$ 
and $\pi^0$ reduce BG from the generic $\tau^+\tau^-$ and $q\overline{q}$ 
continuum: 
$0.6 < \cos \theta_{p \gamma}^{\rm{CM}} < 0.96$ and 
$0.0 < \cos \theta_{p \pi^0}^{\rm{CM}} < 0.95$.  
The opening angle between the two tracks in the CM frame
is also required to be larger than 90$^{\circ}$. 
Constraints on the missing momentum $p_{\rm{miss}}$ and polar-angle 
$\theta_{\rm{miss}}$ are imposed to ensure that the missing particles 
are undetected neutrino(s) 
rather than photons or charged particles that escaped detection:
$p_{\rm{miss}} > 0.6$ GeV/$c$ for $\overline{p}\gamma$ and 
%$p_{\rm{miss}} > 0.5$ GeV/$c$ for the $\overline{p}\pi^0$, and 
$p_{\rm{miss}} > 0.5$ GeV/$c$ for $\overline{p}\pi^0$, and 
$-0.866 < \cos\theta_{\rm{miss}} < 0.956$ for both modes.  
A requirement on the opening angle $\theta_{\rm{tag-miss}}^{\rm{CM}}$ between 
the tag-side track and the missing particle helps to remove 
$\tau^+\tau^-$ events: 
$\cos\theta_{\rm{tag-miss}}^{\rm{CM}} > 0.3$ for $\overline{p}\gamma$ and 
$\cos\theta_{\rm{tag-miss}}^{\rm{CM}} > 0.0$ for $\overline{p}\pi^0$. 

\begin{figure}[t]
\begin{center}
 \resizebox{.3\textwidth}{!}{\includegraphics
 {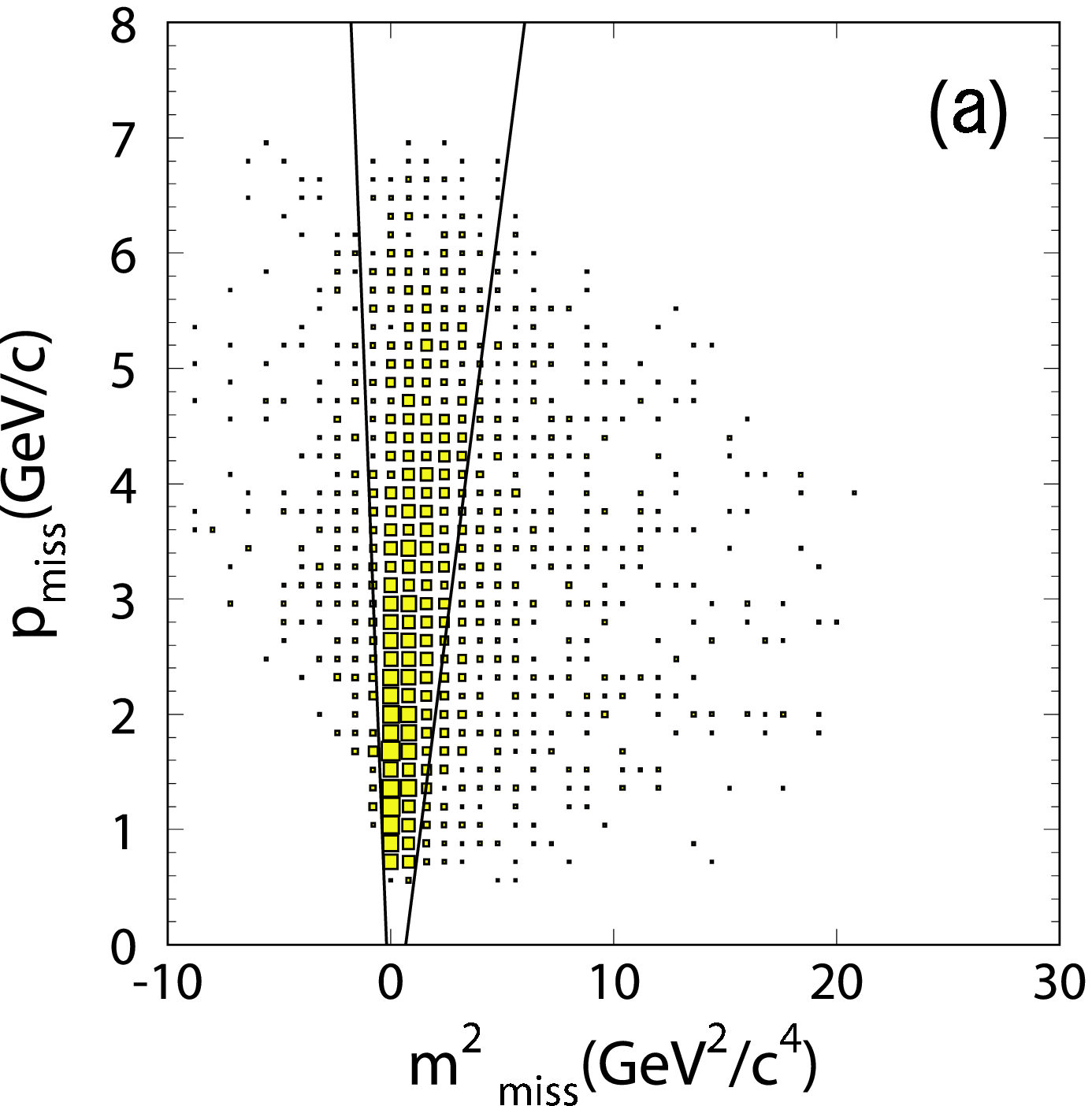}}
 \resizebox{.3\textwidth}{!}{\includegraphics
 {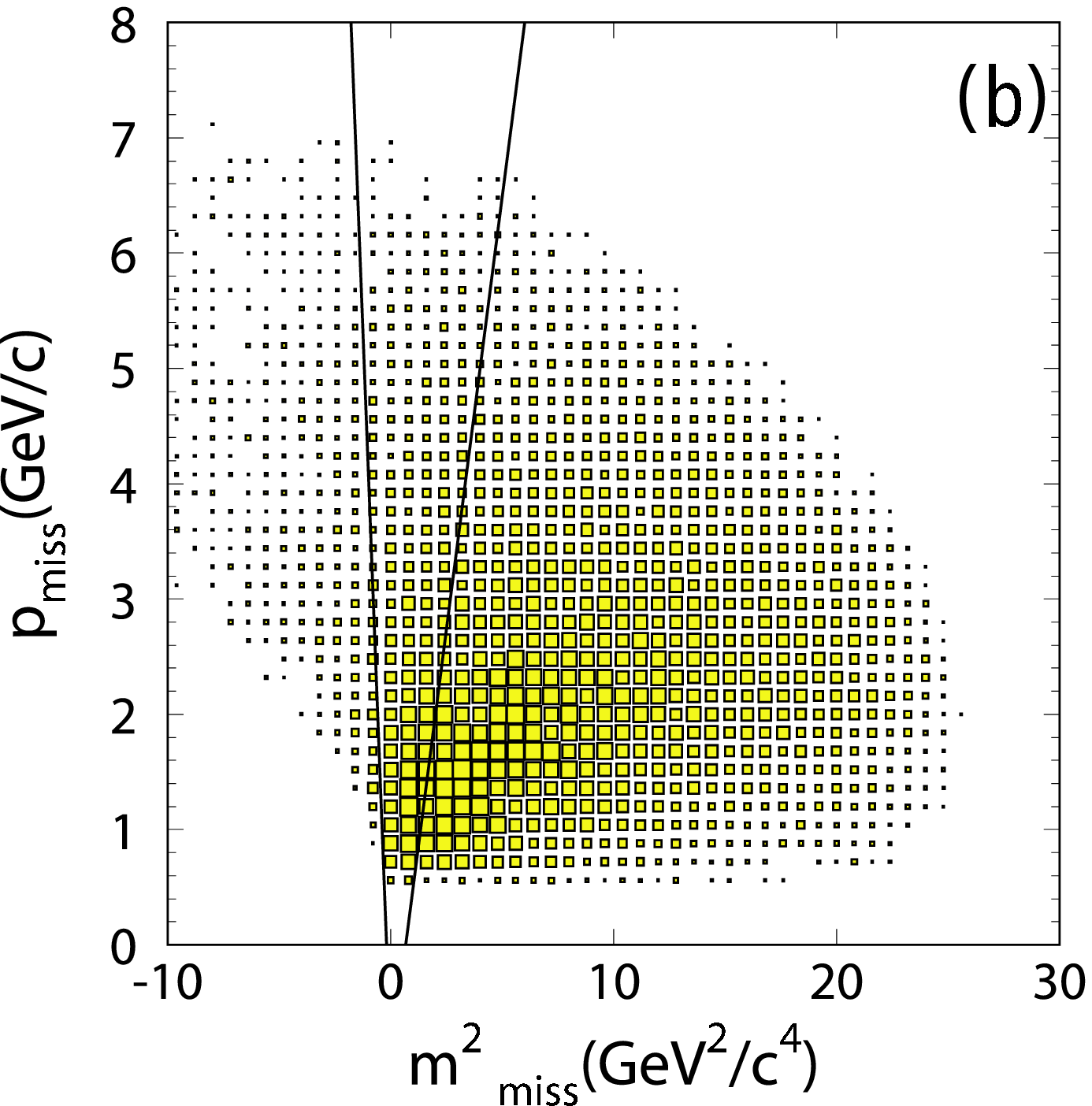}}
 \resizebox{.3\textwidth}{!}{\includegraphics
 {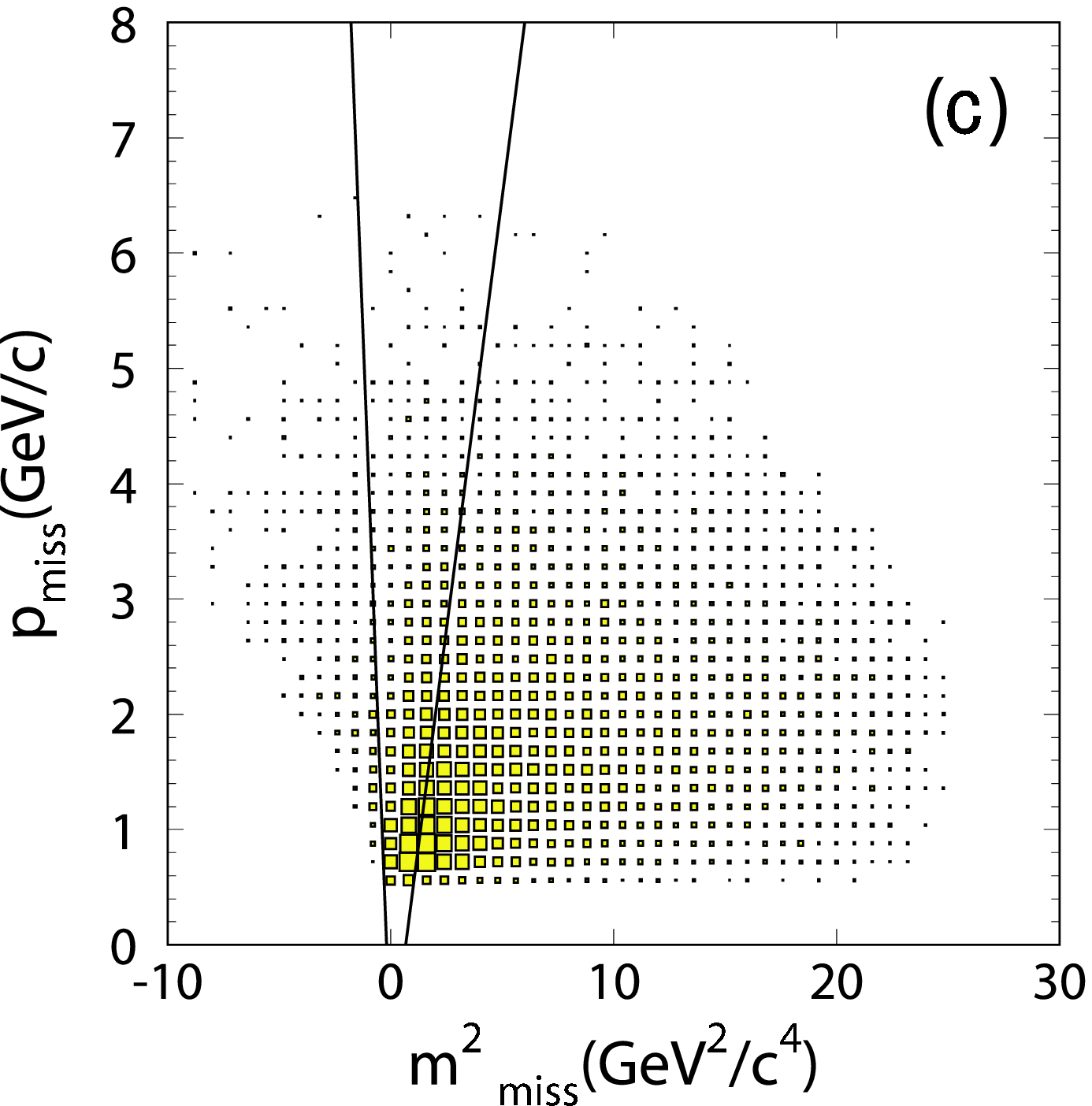}}
 \caption{$p_{\rm{miss}}$ vs $m_{\rm{miss}}^2$ plots
 of $\tau^- \rightarrow \overline{p} \gamma$:  
 (a) signal MC, (b) $\tau^+\tau^-$ MC and (c) $q\overline{q}$ ($uds$) 
continuum MC. The area between the lines is the selected region. 
 \label{fig:pmiss_vs_mmiss2}
 }
 \end{center}
\end{figure}

An additional condition is imposed on the relation between $p_{\rm{miss}}$ and 
the mass squared of a missing particle $m_{\rm{miss}}^2$ 
(see Fig.\ref{fig:pmiss_vs_mmiss2}): 
$p_{\rm{miss}} > -3~ m_{\rm{miss}}^2 - 1$ and $p_{\rm{miss}} > 1.2~ m_{\rm{miss}}^2 - 1$
for $\overline{p}\gamma$, and 
$p_{\rm{miss}} > -0.8~ m_{\rm{miss}}^2 + 0.3$ and $p_{\rm{miss}} > 1.6~ m_{\rm{miss}}^2 - 1$ 
for $\overline{p} \pi^0$.
This cut removes 81\% of the remaining $\tau^+\tau^-$ and 73\% of continuum BG's while 
retaining 78\% of the signal.

\vspace{3 mm}
Particle identification, 
that is very important in this measurement, 
is based on the responses of subdetectors such as	
the ratio of the energy 
deposited in ECL to the momentum measured by CDC, the shower shape 
in ECL, particle range in KLM, the hit information from the 
threshold type ACC, 
$dE/dX$ in CDC and time-of-flight
from TOF.
We use likelihood ratios to distinguish hadron species, 
for instance, 
${\cal{P}}(p/\pi) = {\cal{L}}_p/({\cal{L}}_p + {\cal{L}}_{\pi} )$,
where ${\cal{L}}_i$ is the likelihood for the detector response 
to the track with flavor hypothesis $i$. 
For lepton identification,
we use an electron probability ${\cal P}(e)$ and 
a muon probability ${\cal P}({\mu})$ determined by the detector responses.

We demand ${\cal P}(e) < 0.8$ and ${\cal P}(\mu) < 0.8$ for the signal-side track  
in order to remove Bhabha and $\mu^+\mu^-$ processes. 
To identify protons, the ACC plays an important role
since it allows us
to distinguish flavors by the threshold momentum: 
for protons, this is 
$p_{\rm{p-th}}^{\rm{ACC}}\simeq 5$ GeV/$c$ in the barrel region and 
$\simeq 4$ GeV/$c$ in the forward endcap, while for pions is
$p_{\rm{\pi-th}}^{\rm{ACC}} > 1$ GeV/$c$. 

Most BG tracks are pions with a rather high momentum up to 5~GeV/$c$, and
the momentum of most protons ranges from 1 to 5 GeV/$c$. 
Therefore, we can identify protons and remove pions by requiring ACC 
not to fire at 
1 GeV/$c$ $< p < p_{\rm{p-th}}^{\rm{ACC}}$. 
With these criteria, 90\% of the BG is removed, while 80\% of the signal 
remains.

The information from TOF and CDC is combined with that from the ACC 
into ${\cal{L}}_{p}$ and ${\cal{L}}_{\pi}$ 
to further reduce $\pi$ BG as well as  $K$'s in $\overline{p}\pi^0$. 
The requirement ${\cal P}(p/\pi) > 0.8$ removes more than 
90\% of the BG while retaining 87\% of the signal.
In the $\overline{p}\pi^0$ mode, background processes with kaons,
mostly from $\tau \rightarrow K^{*}\nu$ and 
$K^{*} \rightarrow K \pi^0$ are rejected by demanding 
${\cal P}(p/K) > 0.8$: 70\% of the $K$'s are removed while
77\% of the signal is retained.

\subsection{Expected backgrounds and blind analysis}

A signal candidate is examined in the two-dimensional 
space of the $\overline{p}\gamma$/ $\overline{p}\pi^0$ invariant 
mass, $M_{\rm {inv}}$, and the difference of its energy from the 
beam energy in the CM system, $\Delta E$.
A signal event should have $M_{\rm {inv}}\simeq m_{\tau}$ and 
$\Delta E \simeq 0$. 
The  $M_{\rm {inv}}$ and  $\Delta E$  resolutions  are evaluated 
from the MC distributions around the peak using an asymmetric
Gaussian shape to account for initial state radiation and ECL
energy leakage for photons:  
$\sigma^{\rm{high/ low}}_{M_{\rm {inv}}} = 10.7/16.7$ MeV/$c$$^2$
and 
$\sigma^{\rm{high/ low}}_{\Delta E} = 35.2/62.8$ MeV
for the $\overline{p}\gamma$, and 
$\sigma^{\rm{high/ low}}_{M_{\rm {inv}}} = 11.3/14.9$ MeV/$c$$^2$
and 
$\sigma^{\rm{high/ low}}_{\Delta E} = 34.3/57.1$ MeV
for the $\overline{p}\pi^0$, 
where the ``high/low'' superscript indicates the higher/lower side 
of the peak. 

\begin{figure}[!ht]
\begin{center}
\begin{minipage}{1.05\textwidth}
 \resizebox{0.4\textwidth}{0.4\textwidth}
 {\includegraphics
 {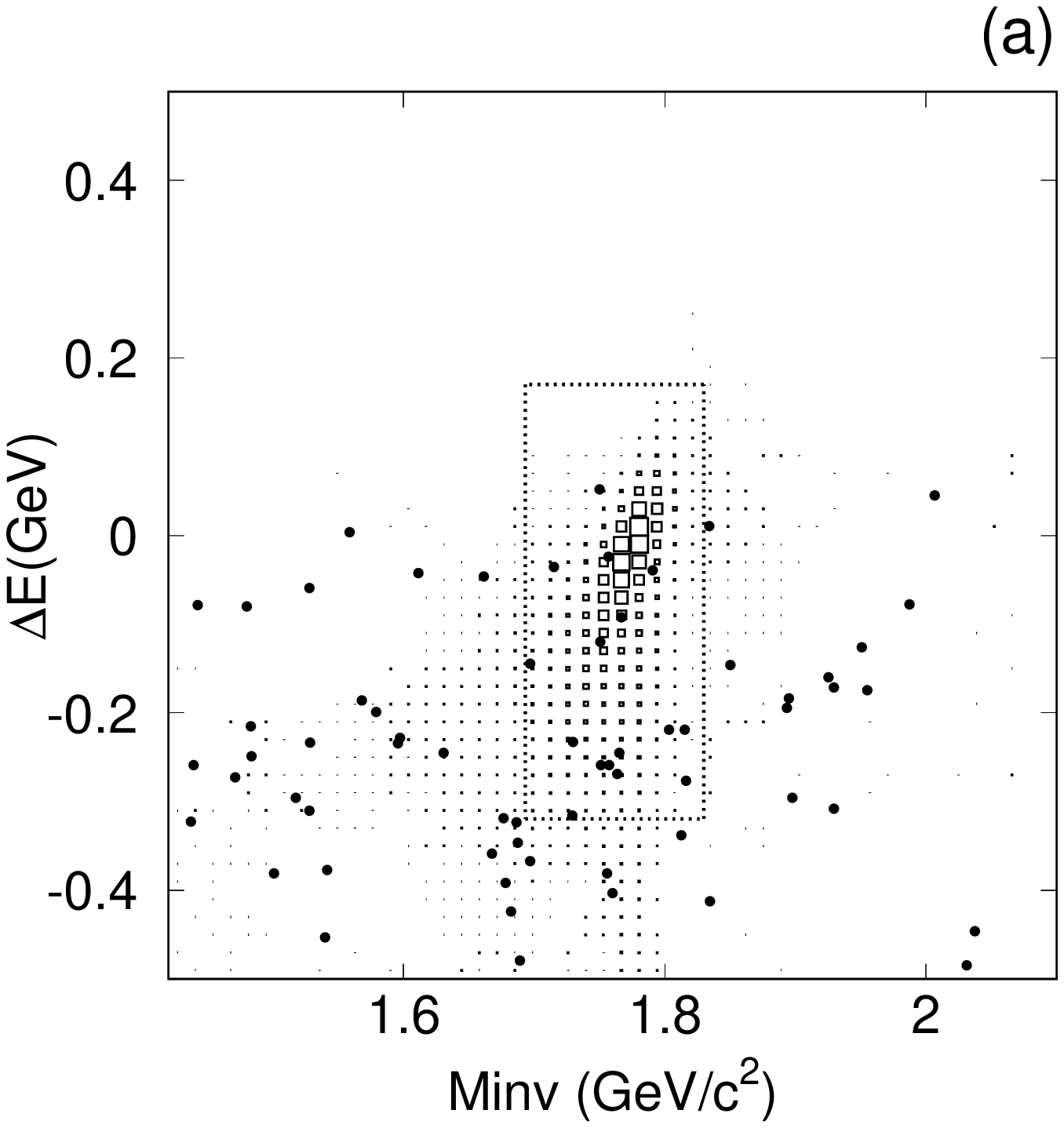}}
 \resizebox{0.4\textwidth}{0.4\textwidth}
 {\includegraphics
 {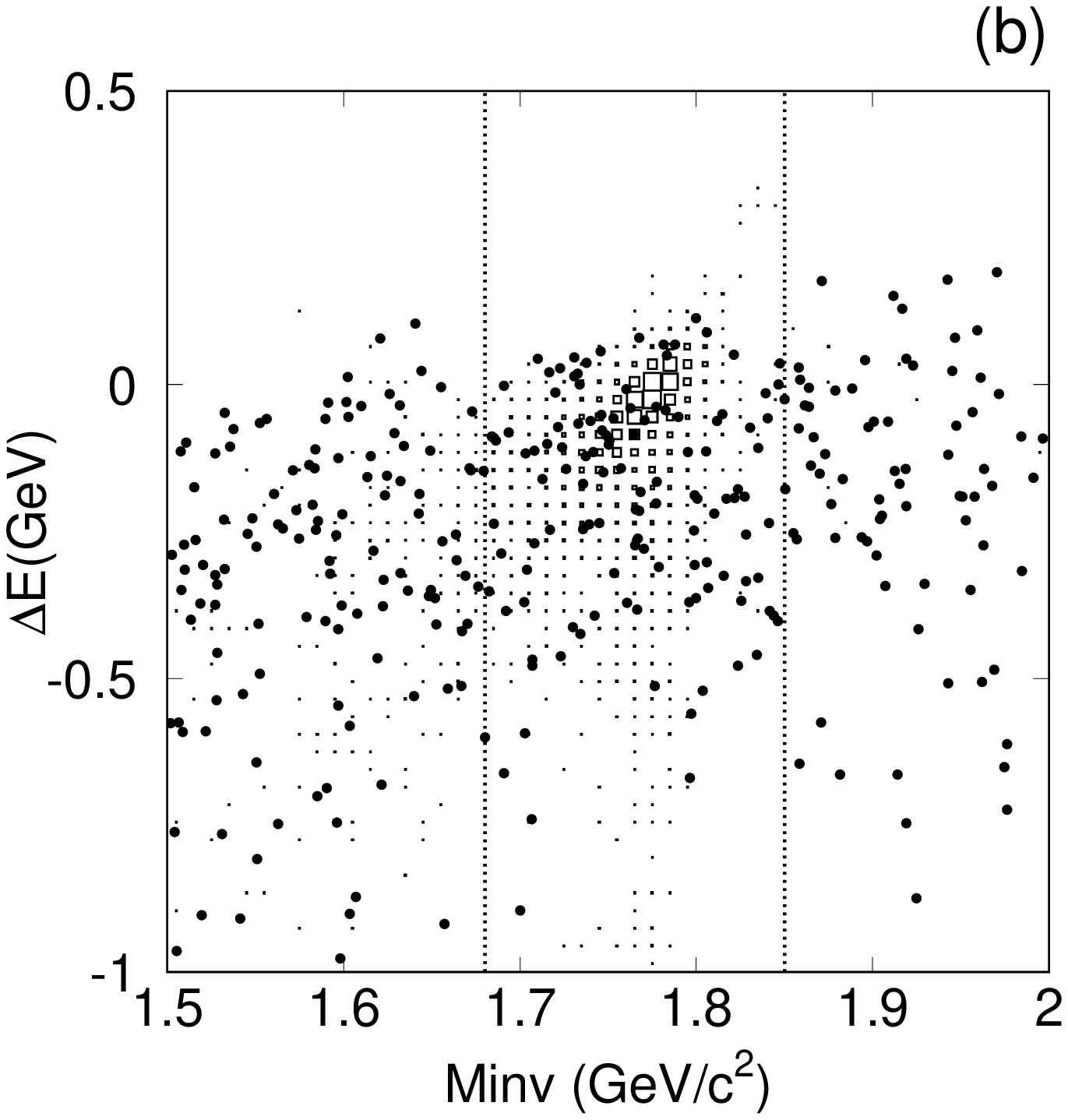}}
\end{minipage}
 \caption{ 
 $\Delta E$ -- $M_{\rm{inv}}$ plot after opening the blind areas 
  (inside the dotted lines) for
 (a) $\overline{p}\gamma$ and (b) $\overline{p}\pi^0$ decay modes. 
% The data are shown by bold dots and the signal MC by small dots.
 The data are shown by bold dots and the signal MC by the boxes.
  }
 \label{fig:open_blind_de_vs_minv}
\end{center}
\end{figure}

To avoid any bias in extracting the result, 
we blind the following region: 
$\pm 5\sigma_{M_{\rm inv}}$ and $\pm 5\sigma_{\Delta E}$ 
for $\overline{p}\gamma$, 
and the 1.68 GeV$/c^2 < {M_{\rm inv}} <$ 1.85 GeV/$c^2$ band region 
for $\overline{p}\pi^0$ (see Fig.~\ref{fig:open_blind_de_vs_minv}). 
Since the $\pm5 \sigma$ region for 
the $\overline{p} \pi^0$ mode contains significantly more events than in the
$\overline{p} \gamma$ case, we blind a wider region for the former mode.

To estimate the expected number of BG events in the blinded region,
we approximate the data distribution
by a combination of Landau and Gaussian functions with a few parameters 
in the region 1.42 GeV/$c^2$ $< M_{\rm inv} < 2.10$ GeV/$c^2$
for $\overline{p}\gamma$,
and 
by an asymmetric Gaussian function with a few parameters 
in the region 1.5 GeV/$c^2$ $< M_{\rm inv} < 2.0$ GeV/$c^2$
for $\overline{p}\pi^0$ 
as shown in
Fig.~\ref{fig:3}.  
One finds in these areas, excluding the blinded region,
49 and 195 data events 
for $\overline{p}\gamma$ and $\overline{p}\pi^0$, respectively, 
and $51.8 \pm 6.1$ and $178.4 \pm 12.8$ corresponding MC events
(properly normalized to the luminosity of the data), 
after application of all cuts.
The number of BG events expected from the BG functions
in the 5$\sigma$ regions is
$9.1 \pm 1.7$ and $52.2 \pm 7.3$
events for 
$\overline{p}\gamma$ and $\overline{p}\pi^0$, respectively. 
The independent evaluation of the number of BG events in the
5$\sigma$ regions based on the MC simulation of the BG processes gave 
6.1 $\pm$ 2.0
and
43.4 $\pm$ 6.6
events for $\overline{p}\gamma$ and $\overline{p}\pi^0$, respectively,
in reasonable agreement with the sideband based evaluation above.
The data and signal MC distributions in $\Delta E$ vs $M_{\rm{inv}}$ 
are shown for both modes in 
Fig.~\ref{fig:open_blind_de_vs_minv}.

The signal detection efficiency for the $\pm 5\sigma$ box is 
evaluated from MC as 9.4\% and 5.8\% for $\overline{p}\gamma$ and 
$\overline{p}\pi^0$, respectively. 

\subsection{Blind opening and evaluation of the branching fractions}

\begin{figure}[!h]
\begin{minipage}{1.05\textwidth}
 \hspace*{-5 mm} \resizebox{0.43\textwidth}{0.8\textwidth}{\includegraphics
 {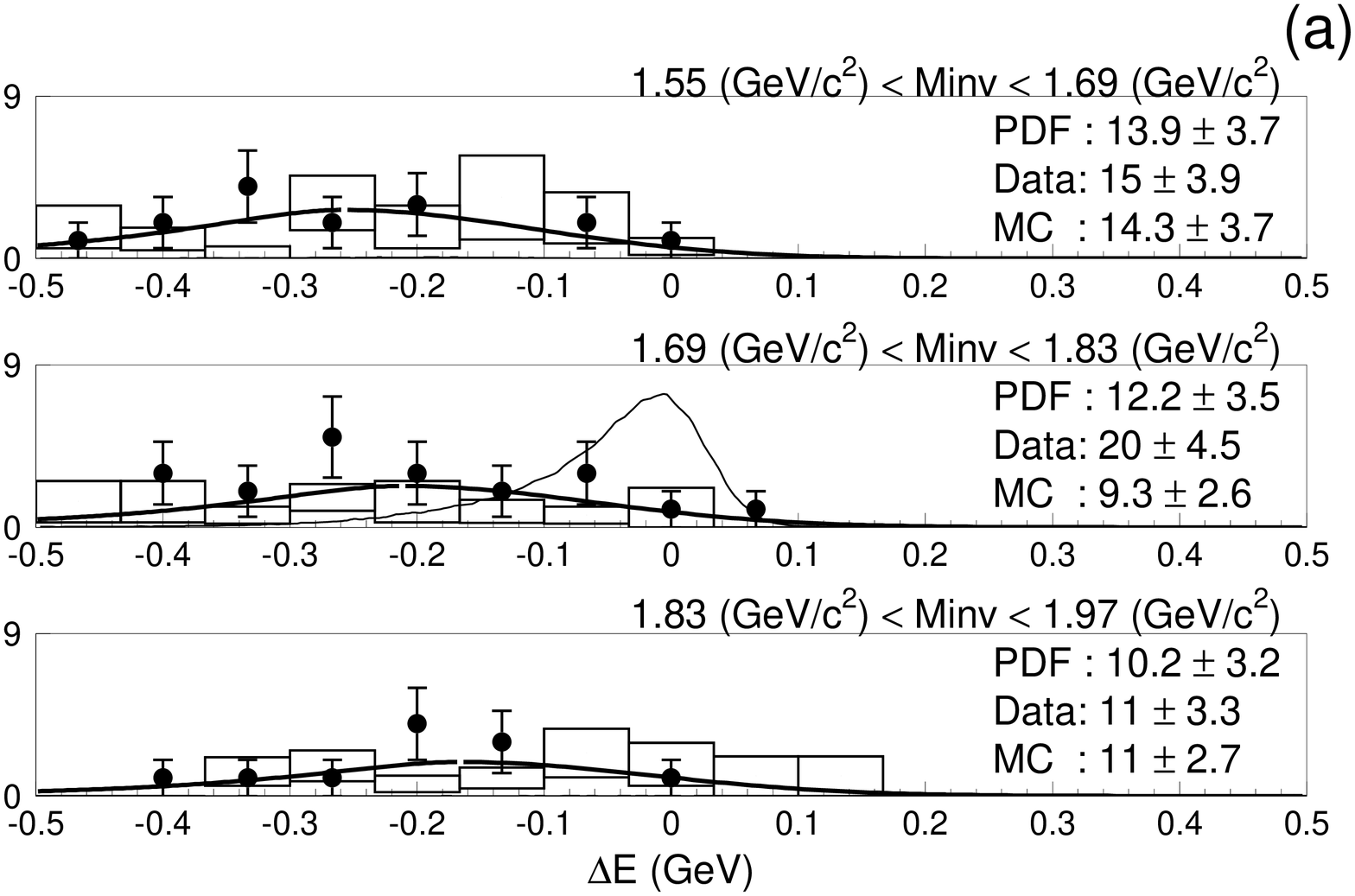}}
 \hspace*{-5 mm} \resizebox{0.55\textwidth}{0.8\textwidth}{\includegraphics
 {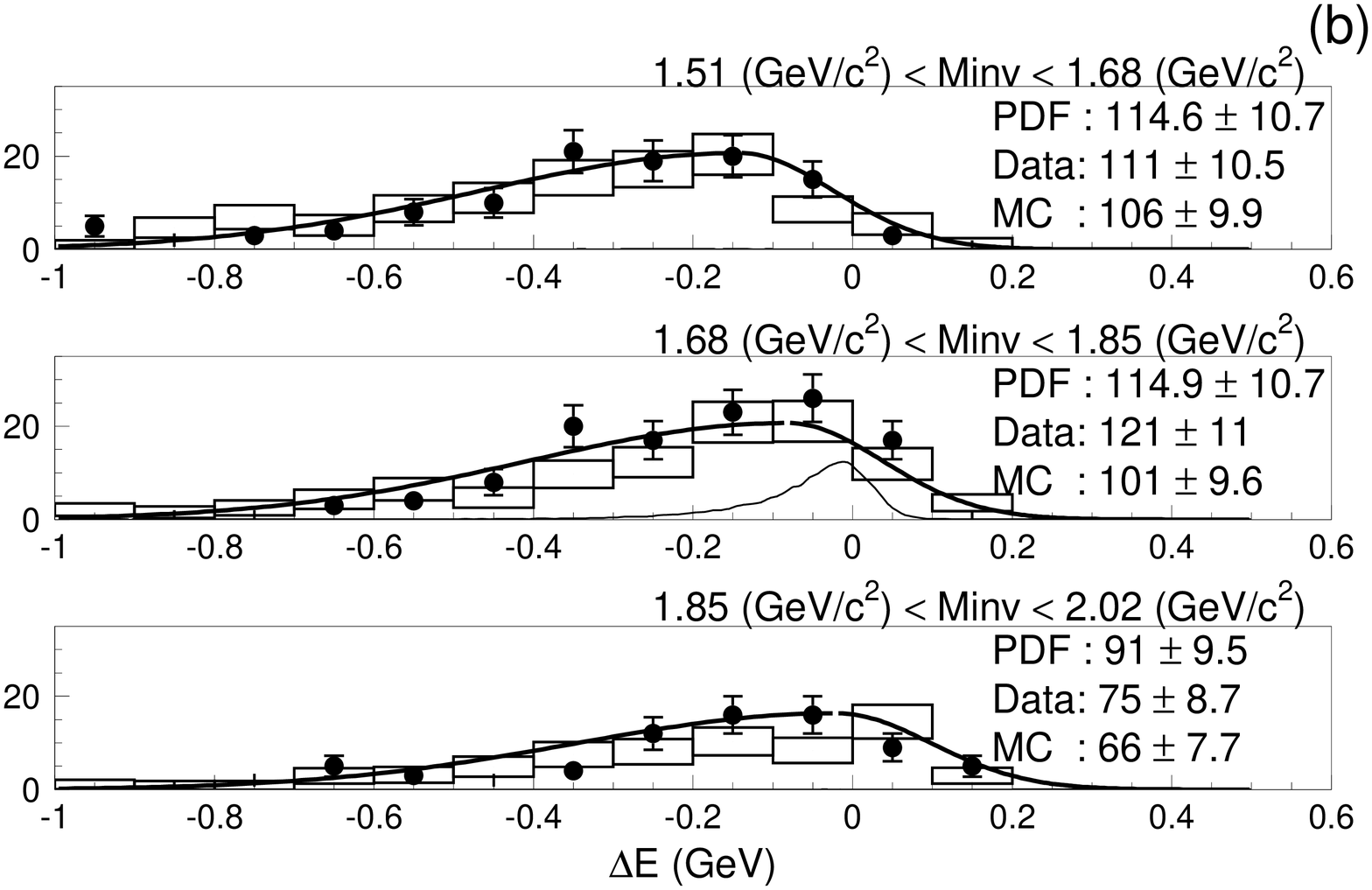}}
\end{minipage}
 \vspace*{-55mm}
 \caption{
The $\Delta{E}$ distributions 
for (a) $\overline{p} \gamma$ and (b) $\overline{p} \pi^0$.
In both (a) and (b) the upper, middle and 
lower figures correspond to the indicated $M_{\rm inv}$ regions:
the middle one is the blind area while the upper and lower correspond
to the sidebands. All three regions have the same width. 
The closed circles with error bars are the data
observed after opening the signal box,
the histogram with error rectangles shows the results 
of MC simulation for the BG processes (generic $\tau^+\tau^-$ and 
continuum $q\bar{q}$). 
The thick solid curve shows the contribution of the
BG parameterized in the sideband regions. 
The expected signal ($\mathcal{B}=10 \times 10^{-7}$) is shown
by the thin solid line. 
The data in the middle figures are blinded until the last stage
of the analysis. 
}
\label{fig:3}
\end{figure}

Data distributions 
after opening the blinded regions
are shown in 
Figs.~\ref{fig:open_blind_de_vs_minv} and \ref{fig:3}. 
We
find
16 and 70 events in the 5$\sigma$ regions
for $\overline{p}\gamma$ and $\overline{p}\pi^0$, respectively,  
while the number of expected BG events is
$9.1 \pm 1.7$ and $52.2 \pm 7.3$
as evaluated above using the data sidebands.
The differences between the number of observed data events and
the BG expectations, 
6.9 $\pm$ 4.3 for $\overline{p}\gamma$
and
17.8 $\pm$ 11.1 for $\overline{p}\pi^0$, 
are not statistically significant.

To extract the number of signal events,
we applied an unbinned extended maximum likelihood method (UEML),
which is more sensitive than the binned one since it 
uses the complete information about events. In this method the likelihood 
function is defined as
\begin{equation}
  {\cal L}(s,b) = \frac{e^{-(s+b)}}{N!} \prod_{i=1}^{N} (s S_i + b B_i), 
\label{eq:likelihood}
\end{equation} 
where $N$ is the number of the observed events, $s$ and $b$ are 
the free parameters corresponding to
the number of signal and background events, respectively,  
and $S_i$ and $B_i$ are the values of the probability density 
functions of the signal and BG for the $i$-th event. 
$S_i$ is given by the signal MC distributions, and $B_i$ is the 
BG function obtained above and normalized to unity. 

The maximum likelihood fit for the $\pm 5\sigma$ region yields 
$s_0$ = 0.16 and $b_0$ = 15.84 for $\overline{p}\gamma$ and 
$s_0$ = 3.09 and $b_0$ = 66.91 for $\overline{p}\pi^0$, respectively.
%Following~\cite{cite:Narsky_1999kt},
Following Ref.~\cite{cite:Narsky_1999kt},
the upper limit at the 90\% confidence level (C.L.) is obtained 
by means of toy MC, as described below.
For every assumed expected signal yield $\tilde{s}$,
10,000 samples are generated, for each of which the number of signal 
and BG events is determined by Poisson statistics with the mean 
values $\tilde{s}$ and $b_0$, respectively. 
We then assign the $\Delta E$ and $M_{\rm{inv}}$ values to these 
events according to their distributions; a UEML fit is performed 
for every prepared sample 
to extract the signal yield ($\tilde{s}_0$); 
the confidence level for an assumed $\tilde{s}$ is defined as the 
fraction of the samples whose $\tilde{s}_0$ exceeds $\tilde{s}$. 
This procedure is repeated until we find
the value of $\tilde{s}_{90}$
that gives a 90\% chance
of $\tilde{s}_0$ being larger than $\tilde{s}$. 
The resulting values are 
$\tilde{s}_{90}$ = 3.7 events for $\overline{p}\gamma$
and $\tilde{s}_{90}$ = 9.8 events for $\overline{p}\pi^0$. 

The upper limit on the branching fraction $\mathcal{B}$ at the 90\% C.L.
is 
then calculated as 
\begin{equation}
{\cal B} < \frac{\tilde{s}_{90}}{2 \varepsilon N_{\tau\tau}}, 
  \label{eq:branching_fraction}
\end{equation}
where the number of produced $\tau$-pairs is $N_{\tau\tau}$ =  
$78.9 \times 10^6$ 
and
$140.0 \times 10^6$
for $\overline{p}\gamma$ and 
$\overline{p}\pi^0$, respectively, 
and the detection efficiencies are $\varepsilon$ = 9.4\% and 5.8\%. 
The resulting upper limits are  
${\cal B}(\tau^- \rightarrow \overline{p} \gamma) < 2.5 \times 10^{-7}$  and 
${\cal B}(\tau^- \rightarrow \overline{p} \pi^0) < 6.1 \times 10^{-7}$ at 90\% C.L.

Systematic uncertainties related to the detector sensitivity 
$2 \varepsilon N_{\tau\tau}$ in the denominator of 
%eq.(\ref{eq:branching_fraction}) are evaluated to be 5.9\% for 
Eq. (\ref{eq:branching_fraction}) are evaluated to be 5.9\% for 
$\overline{p}\gamma$ and 8.7\% for $\overline{p}\pi^0$. 
The individual contributions to the uncertainties
for $\overline{p}\gamma$ ($\overline{p}\pi^0$) are 
2.0\%(2.0\%) from tracking efficiency, 
3.0\%(6.0\%) from photon reconstruction efficiency, 
2.0\%(4.0\%) from selection criteria, 
3.0\%(3.0\%) from trigger efficiency, 
2.5\%(3.0\%) from proton identification, 
0.2\%(0.3\%) from MC statistics, 
1.4\%(1.4\%) from luminosity evaluation, 
and 
%0.03\%(0.03\%) from the $\tau^+\tau^-$ cross-section,
%respectively.  
0.03\%(0.03\%) from the $\tau^+\tau^-$ cross-section.

The systematic uncertainty in $\tilde{s}_{90}$ is estimated by varying 
the parameters of the BG functions by $\pm 1\sigma$. 
This gives uncertainties of $\pm 0.77$ events for $\overline{p}\gamma$ and 
$\pm 0.75$ events for 
$\overline{p}\pi^0$.
The upper limits on the branching fractions taking into account
all the systematic uncertainties are calculated to be 
\begin{eqnarray}
 {\cal B}(\tau^- \rightarrow \overline{p}\gamma) &<& 3.0 \times 10^{-7} \\
 {\cal B}(\tau^- \rightarrow \overline{p}\pi^0) &<& 6.5 \times 10^{-7}
\end{eqnarray}
at 90\% C.L. with 86.7 fb$^{-1}$ and 153.8 fb$^{-1}$ of data, 
respectively. 

The $\overline{p} \pi^0$ mode
should be discussed briefly
since
some excess of events in the signal area can be seen in the
$\Delta E$ distribution
of
Fig. \ref{fig:3}.  
%From the s dependence of the likelihood function, 
%a 68\% confidence interval for s0 is $-1.6 < s_0 < 8.8$; 
From the $s$ dependence of the likelihood function, 
a 68\% confidence interval for $s_0$ is $-1.6 < s_0 < 8.8$; 
in other words, $s_0$ is consistent with 0 within 1 $\sigma$.
If we use toy MC, a probability
to have $s_0 > 3.09$ when the signal yield $\tilde{s}=0$ is 19\%.
We conclude that the observed excess can be due a statistical fluctuation.

\section{Analysis of $\bm{\tau^-\rightarrow \overline{\Lambda}\pi^-}$ and $\bm{\Lambda \pi^-}$}

\subsection{Event Selection}

%The experimental signature these events has one $\tau$ lepton (signal side) decaying to 
The experimental signature of these events has one $\tau$ lepton (signal side) decaying to 
$\Lambda\pi$, $\Lambda$ to $p\pi$ and the other (tag side) 
decaying via a 1-prong mode: 
\begin{center}
$\left\{\tau^- \rightarrow (\overline{p}\pi^+) + \pi^- \right\} ~+
 ~ \left\{ \tau^+ \rightarrow ({\rm a~track})^+ + (n^{\rm TAG}_{\gamma} > 0)
 + X(\rm{missing}) \right\}$.
\end{center}
We consider here the $B-L$ conserving decay modes: 
$\tau^-\rightarrow\overline{\Lambda}\pi^-$ with 
$\overline{\Lambda}\rightarrow\overline{p}\pi^+$
and 
$\tau^+\rightarrow\Lambda\pi^+$ with 
$\Lambda\rightarrow p\pi^-$. 
We also consider the $B-L$ violating decay modes:
$\tau^-\rightarrow \Lambda \pi^-$ with 
$\Lambda \rightarrow p\pi^-$ and 
$\tau^+ \rightarrow \overline{\Lambda} \pi^+$
with
$\overline{\Lambda }\rightarrow \overline{p}\pi^+$.
The experimental signature for these modes is:
\begin{center}
$\left\{\tau^- \rightarrow (p\pi^-) + \pi^- \right\} ~+~ 
 \left\{ \tau^+ \rightarrow ({\rm a~track})^+ + (n^{\rm TAG}_{\gamma} > 0)
 + X(\rm{missing}) \right\}$.
\end{center}
We denote the pion from $\tau\rightarrow\Lambda\pi$ as $\pi_1$ and 
the pion from $\Lambda\rightarrow p\pi$ as $\pi_2$. 
We can distinguish between the $B-L$ conserving and violating modes
by the charge of these pions:
the $B-L$ conserving decay modes have an opposite sign combination
on the $\pi_1$ and $\pi_2$ charges, 
while the $B-L$ violating modes have a same sign combination. 
As in the case of $\tau^- \rightarrow \overline{p}\gamma$
and $\overline{p}\pi^0$ modes, 
tracks and photons should have 
$p_t > 0.1$ GeV/$c$ 
and $E_{\gamma} > 0.1$ GeV, respectively, with a polar angle
satisfying $-0.866 < \cos\theta < 0.956$. 

We first demand that the four tracks have a net zero charge.  
The magnitude of the thrust is required to be larger than 0.9 to suppress
the $q\overline{q}$ continuum background. 
The event should have a 1-3 prong configuration relative to the 
plane perpendicular to the thrust axis. 
We select $\overline{\Lambda}$ candidates via the $\overline{p}\pi^+$ 
decay channel based on the angular 
difference between the $\overline{\Lambda}$ flight direction
and the direction pointing from 
the interaction point to the decay vertex
(see, Ref.\cite{Lambda_rec} for more details). 
The proton from the $\overline{\Lambda}$ decay is identified by demanding 
${\cal{P}}(\overline{p}/\pi) > 0.6$. 
In order to avoid fake $\overline{\Lambda}$ candidates in which two tracks
in the signal side are an $e^-e^+$ pair from a photon conversion,
the electron veto is imposed on the three tracks in the signal side.
The reconstructed $\overline{\Lambda}$ candidate mass should be
within $\pm5$ MeV/$c$$^2$ of the nominal $\Lambda$ mass and
$p^{\rm{CM}}_{\Lambda} > 1.75$ GeV/$c$ is required
to reduce contributions from the generic $\tau^+\tau^-$ and $q\overline{q}$ continuum 
as shown in Fig.~4.       
\begin{figure}[h]
\begin{center}
 \resizebox{0.45\textwidth}{0.35\textwidth}{\includegraphics
 {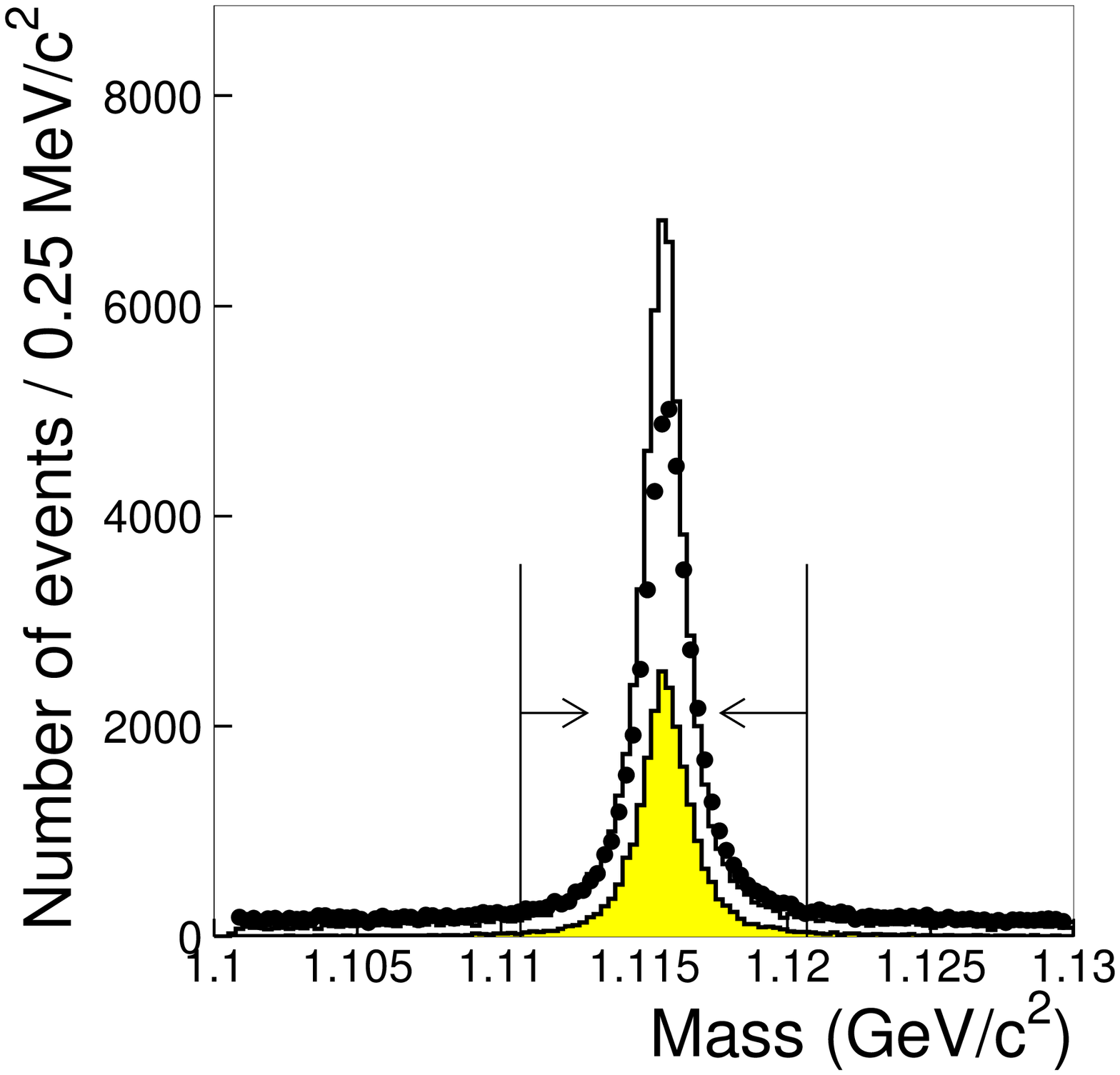}}
 \resizebox{0.45\textwidth}{0.35\textwidth}{\includegraphics
 {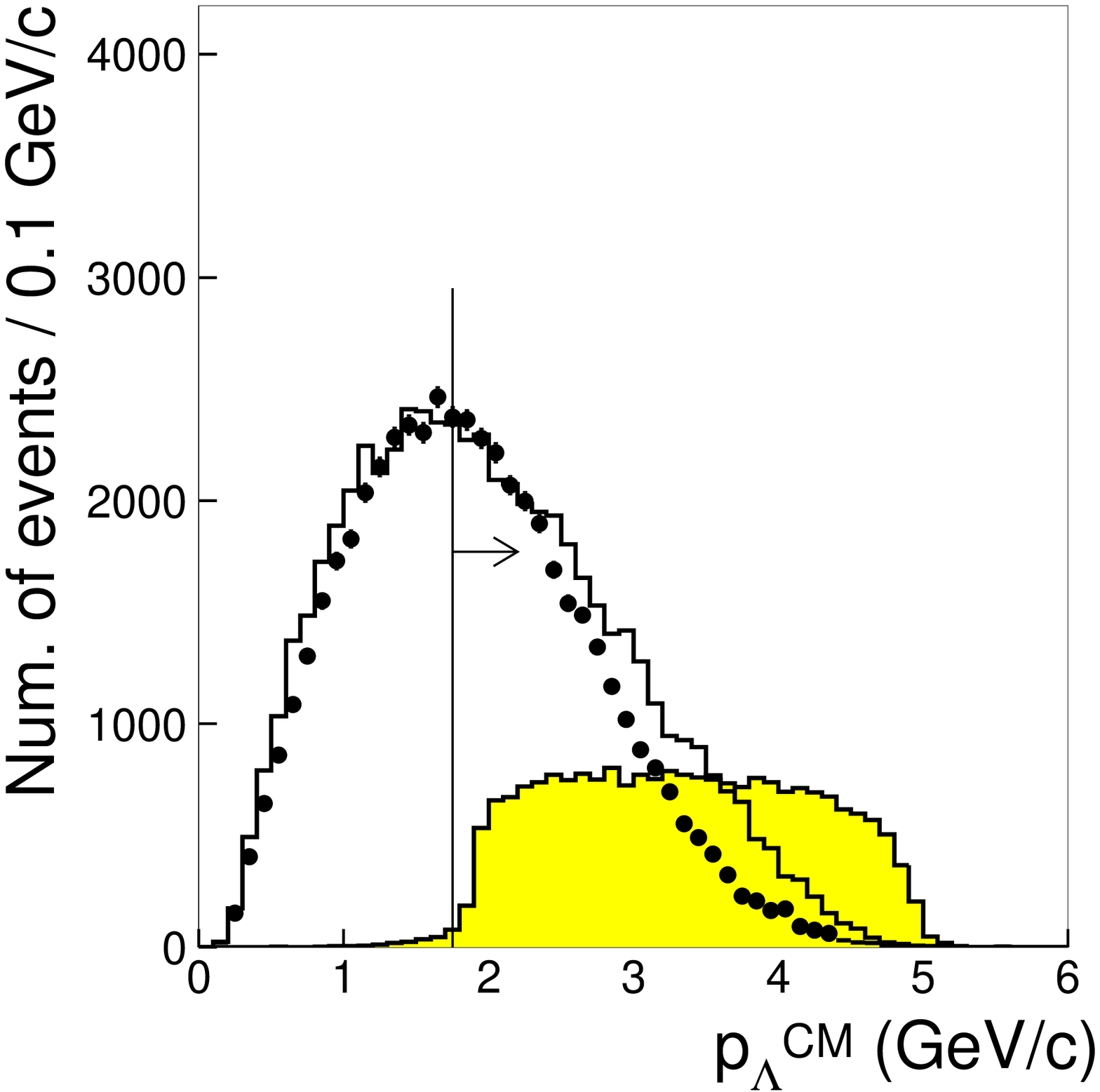}}
 \caption{ Reconstructed $\overline{\Lambda}$ candidate mass (left) 
 and momentum (right) in the CM frame.
 The signal MC distributions are indicated by the filled histogram, 
 all BG's including $\tau^+\tau^-$ and $q\overline{q}$ by the open histogram, 
 and closed circles are data.
 While the signal MC is normalized arbitrarily, 
 the data and MC are normalized to the same luminosity.
 The selected area is indicated by the lines with arrows. 
}
\label{fig:4}
\end{center}
\end{figure}

As in the $\overline{p}\gamma$ and $\overline{p}\pi^0$ cases, 
the following criteria are imposed: 
$5.29 < E^{\rm{CM}}_{\rm{vis}} < 10.5$ GeV to reject Bhabha, 
two-photon and $\mu^+\mu^-$ reactions;  
$p_{\rm{miss}} > 0.4$ GeV/$c$ and 
$-0.866 < \cos\theta_{\rm{miss}} < 0.956$  
to ensure that a missing particle is a neutrino(s); 
$\cos\theta^{\rm{CM}}_{\rm{miss-tag}} > 0$ to include the 
missing particle in the tag side; 
$n_{\gamma}^{\rm{SIG}}\leq 1$ and $n_{\gamma}^{\rm{TAG}}\leq 2$
to suppress the continuum background.

Both the proton veto
${\cal P}(\overline{p}/\pi) < 0.6$
and kaon veto
${\cal P}(K/\pi) < 0.6$
are 
applied to $\pi_1$
and the tag-side track.

The correlation between the missing momentum $p_{\rm{miss}}$ and 
mass-squared $m^2_{\rm{miss}}$ is considered to further suppress BG's from
generic $\tau^+\tau^-$ and continuum BG: 
$p_{\rm{miss}} > 1.5\times m^2_{\rm{miss}}- 1.0$. 

\subsection{Signal resolutions and blind analysis}

The $M_{\rm inv}$ and $\Delta{E}$ resolutions are evaluated from MC: 
$\sigma^{\rm{high/ low}}_{M_{\rm{inv}}} = 4.6/ 4.0$ MeV/$c$$^2$ and 
$\sigma^{\rm{high/ low}}_{\Delta E} = 22/ 29$ MeV. 
In $\overline{\Lambda}\pi^-$ ($\Lambda \pi^-$), there is no $M_{\rm{inv}}$ tail
due to energy 
leakage from ECL because there are no photons in the final state of this mode.

\begin{figure}[!b]
\begin{center}
 \resizebox{\textwidth}{0.8\textwidth}{\includegraphics
  {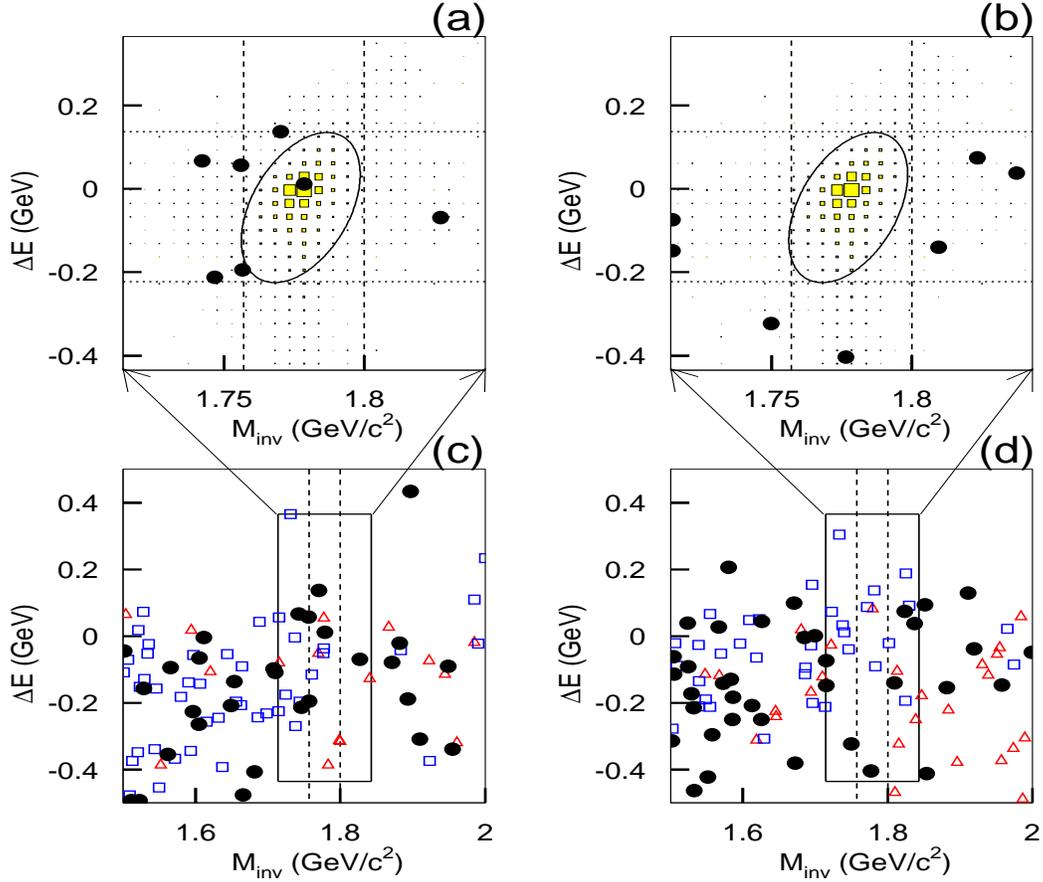}}
 \caption{
Scatter-plot of data and MC events in the $M_{\rm inv}$ -- $\Delta{E}$
plane: (a) and (b) correspond to
the $\pm 15 \sigma$ area for
the $B-L$ conserving and violating modes, respectively,
while (c) and (d) show the region 1.5 $< M_{\rm inv} <$ 2.0 GeV/$c^2$ and
$|\Delta{E}| < 0.5$ GeV for the same modes. The areas of (a) and (b) 
are also shown as solid rectangles in (c) and (d). The areas inside
the dotted lines in (c) and (d) denote the blind signal regions.
The 90\% elliptical region shown by a solid curve in (a) and (b) 
is used for evaluating the signal yield.
In (a) and (b) the vertical dotted lines denote the blind regions
similar to those in (c) and (d);
the regions inside the horizontal dotted lines and outside the vertical dotted lines
are sidebands 
used to estimate the expected BG in the elliptical region.
Closed circles correspond to the data (154 fb$^{-1}$), 
open squares are generic MC $\tau^+\tau^-$  events 
(equivalent luminosity of 297 fb$^{-1}$) and
open triangles are MC $uds$ continuum events
(equivalent luminosity of 104 fb$^{-1}$).
Filled boxes show the MC signal distribution
with arbitrary normalization.
\label{fig:5}
}
\end{center}
\end{figure}

We blind a region over $\pm 5\sigma_{\rm{M_{inv}}}$ and 
$-0.5 < \Delta E < 0.5$ GeV. 
Fig. \ref{fig:5} shows scatter-plots for data and MC samples over $\pm 15\sigma$ 
in the $M_{\rm{inv}}-\Delta E$ plane: 
the number of data and MC events
outside the blinded region  
(bounded by the vertical dotted line in Fig.5 (a) and (b))
are 
$11$ and $13.2\pm 3.5$ events, respectively. 
Good agreement is observed.
The surviving
BG events are due to generic $\tau^+\tau^-$ decays (about 1/2) and
$uds$ continuum
(about 1/2). The former events are dominated by
the $\tau \to a_1(1260) \nu_{\tau}$ decays, in which three
charged pions from the $a_1(1260)$ decay form a fake $\Lambda$ candidate.
The continuum BG events have one true $\Lambda$ which forms a
signal candidate together with another track.

\subsection{Blind opening and evaluation of the branching fraction}

Since there are fewer remaining events compared to 
$\overline{p}\gamma$ and 
$\overline{p}\pi^0$, 
we apply the Frequentist approach using the Feldman \& Cousins 
method~\cite{cite:FC}
rather than the maximum likelihood method. 
We take an elliptical region that contains 90\% of MC signal events 
passing all cuts 
as a signal region, as shown in Fig.~\ref{fig:5} (a) and (b).
It results in the signal detection efficiency $\varepsilon = 11.8\%$
for the $B-L$ conserving and $\varepsilon = 11.7\%$ for the $B-L$ violating modes,
respectively.

From Fig.~\ref{fig:5} (c) and (d)
we assume the BG distribution to be flat along the
$M_{\rm{inv}}$ axis, and then obtain the expected BG in the ellipse 
as $1.7 \pm 0.8$ events for each of the two modes,
using sideband regions, 
inside the horizontal dotted lines and outside the vertical dotted lines,
as shown in the Fig. 5 (a) and (b).
We open the blinded region and find
only one data event in the ellipse for the $B-L$ conserving mode
and 
no data events for the $B-L$ violating mode 
(see Fig.~\ref{fig:5} (a) and (b), respectively). 
The upper limits on the signal yields at 90\% C.L. 
are obtained by the Feldman-Cousins method as 
$s_{90} = 2.8$ 
and  
$s_{90} = 1.2$, 
respectively. 
The upper limits on the branching fraction are then calculated as
\begin{equation}
{\cal B}(\tau \rightarrow \Lambda \pi) 
<  \frac{s_{90}}{2 \varepsilon N_{\tau\tau}{\cal B}
( \Lambda \rightarrow p \pi)}
\end{equation}
where $N_{\tau\tau} = 140.0\times 10^6$ and 
${\cal B}(\Lambda \rightarrow p \pi) = 0.639$~\cite{PDG}. 
The resulting values are
${\cal B}(\tau^-\rightarrow\overline{\Lambda}\pi^-) < 1.3\times 10^{-7}$
and 
${\cal B}(\tau^-\rightarrow \Lambda\pi^-) < 0.58 \times 10^{-7}$. 

Among systematic uncertainties on the detection sensitivity 
$S_0 = 2\varepsilon N_{\tau\tau}{\cal B}(\Lambda\rightarrow p\pi)$, 
$\overline{\Lambda}$ selection and the proton identification 
in $\overline{\Lambda}$ decay contribute 6.0\% and 3.0\%, respectively; 
${\cal{B}}(\Lambda \rightarrow p\pi)$ has an uncertainty of 0.8\% 
\cite{PDG}; 
the trigger efficiency (0.5\%), tracking (4.2\%), selection criteria (4.0\%), 
MC statistics (0.7\%), luminosity (1.4\%) and the $\tau^+\tau^-$ cross-section (0.03\%) 
are also considered. 
All these uncertainties are added in quadrature to 
$9.1\%$ in total.  

The upper limits on the branching fractions at the 90\% C.L.
including systematic errors are then calculated 
by the POLE program~\cite{cite:pole}.
The resulting branching fractions are
\begin{eqnarray*}
&&{\cal B}(\tau^-\rightarrow \overline{\Lambda}\pi^-) < 1.3 \times 10^{-7} \\
&&{\cal B}(\tau^-\rightarrow \Lambda \pi^-) < 0.70 \times 10^{-7}
\end{eqnarray*}
at the 90\% C.L. 

\section{Results}

We obtain the following preliminary upper limits on
the branching fractions:
${\cal{B}}(\tau^-\rightarrow \overline{p}\gamma) < 3.0\times 10^{-7}$, 
${\cal{B}}(\tau^-\rightarrow \overline{p}\pi^0) < 6.5\times 10^{-7}$, 
${\cal{B}}(\tau^-\rightarrow \overline{\Lambda}\pi^-) < 1.3\times 10^{-7}$ 
and 
${\cal{B}}(\tau^-\rightarrow \Lambda \pi^-) < 0.70\times 10^{-7}$ 
at the 90\% confidence level. 

For the latter two modes this is the first search ever performed.  

The resulting upper limits on the branching fraction for 
$\tau^-\rightarrow\overline{p}\gamma$ and $\tau^-\rightarrow\overline{p}\pi^0$ 
improve upon the previous measurements by a factor of 12 and
23, respectively. 
This large improvement is mostly due to the powerful proton identification 
ability of the Belle detector that removes spurious combinatorial BG's as well 
as
higher statistics compared to the previous experiment.

\section*{Acknowledgments}
%***** Acknowledgments *****
% Please paste this acknowledgement into your latex file. 
%----------- Long version, for most papers ----------- 
We thank the KEKB group for the excellent operation of the
accelerator, the KEK Cryogenics group for the efficient
operation of the solenoid, and the KEK computer group and
the National Institute of Informatics for valuable computing
and Super-SINET network support. We acknowledge support from
the Ministry of Education, Culture, Sports, Science, and
Technology of Japan and the Japan Society for the Promotion
of Science; the Australian Research Council and the
Australian Department of Education, Science and Training;
the National Science Foundation of China under contract
No.~10175071; the Department of Science and Technology of
India; the BK21 program of the Ministry of Education of
Korea and the CHEP SRC program of the Korea Science and
Engineering Foundation; the Polish State Committee for
Scientific Research under contract No.~2P03B 01324; the
Ministry of Science and Technology of the Russian
Federation; the Ministry of Education, Science and Sport of
the Republic of Slovenia; the National Science Council and
the Ministry of Education of Taiwan; and the U.S.\
Department of Energy.

\end{document}